# Complex 3-Dimensional Microscale Structures for Quantum Sensing Applications

Brian W. Blankenship[1,2†,] Zachary Jones[2,3†], Naichen Zhao[1], Harpreet Singh[2], Adrisha Sarkar[2], Runxuan Li[1], Erin Suh[1], Alan Chen[1], Costas Grigoropoulos[1*], Ashok Ajoy[2,4*]

1. Laser Thermal Laboratory, Department of Mechanical Engineering, University of California, Berkeley, CA 94720, USA
2. Department of Chemistry, University of California, Berkeley, CA 94720
3. Advanced Biofuels and Bioproducts Process Development Unit, E. O. Lawrence Berkeley National Laboratory, Berkeley, CA 94720, USA.
4. Chemical Sciences Division, Lawrence Berkeley National Laboratory, Berkeley, CA 94720, USA

† These Authors contributed equally
*Corresponding Author

Email: ashokaj@berkeley.edu, cgrigoro@berkeley.edu

**Abstract**
We present a novel method for fabricating highly customizable three-dimensional structures hosting quantum sensors based on Nitrogen Vacancy (NV) centers using two-photon polymerization. This approach overcomes challenges associated with structuring traditional single-crystal quantum sensing platforms and enables the creation of complex, fully three-dimensional, sensor assemblies with sub-microscale resolutions (down to 400 nm) and large fields of view (>1 mm). By embedding NV center-containing nanoparticles in exemplary structures, we demonstrate high sensitivity optical sensing of temperature and magnetic fields at the microscale. Our work showcases the potential for integrating quantum sensors with advanced manufacturing techniques, facilitating the incorporation of sensors into existing microfluidic and electronic platforms, and opening new avenues for widespread utilization of quantum sensors in various applications.

## Introduction

Quantum sensing is an emerging technology that has enabled us to measure and observe the world around us at increasingly miniaturized scales.[1] Quantum sensing approaches often use crystalline defects in wide-bandgap semiconductors, most notably nitrogen vacancy (NV) centers in diamond, based on their ability to host optically addressable spins with unique, spin-state selective photostable fluorescence and long spin coherence times.[2–5] NV-based sensors can yield highly precise measurements of magnetic moments[6–8], electric fields[9,10], strain[11] and temperature[1,12,13] with nanoscale spatial resolution. The design of sensors with useful architectures requires the combination of well controlled placement of NV centers as well as the ability to interrogate the quantum states of the sensors with high sensitivity. In practice the field has largely converged on two general approaches for deploying sensors- each with inherent limitations for structuring NV centers into useful configurations.[14]

One prevalent approach in the field involves modifying single crystalline diamond substrates through either etching processes, which create high surface area and high aspect ratio 2.5D features on the substrate's surface[15,16], or by directly implanting spin defects in 2D patterns near the surface of the substrate.[17–21] However, these methodologies present challenges when attempting to create three-dimensional features like channels and overhangs.

Another approach utilizes nanodiamonds that contain high concentrations of NV centers, which are particularly attractive for intracellular sensing.[13,22] Nonetheless, the techniques for fixing particles in space and creating designer sensor assemblies are still underdeveloped. Although intentional placement of nanodiamond particles can be achieved through optical trapping[8,23] and the creation of permanent structures via complex particle self-assembly processes[24,25], these techniques are limited in terms of the design freedoms inherent to these processes.

Herein we demonstrate an approach to "*3-D print*" complex microscale polymeric structures with embedded nanodiamonds to achieve three-dimensional structuring of quantum sensors. Our goals are to demonstrate a fast, large-area fully 3D approach for arranging and positioning quantum sensors with sub-micrometer scale resolution, ultimately applying them in sensing applications.

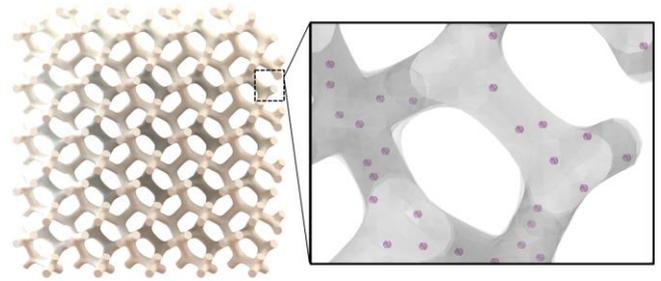

**Figure 1: Motivation and Strategy (Left)** Complex microscale porous structures as shown can enable new sensing capabilities but are impractical to fabricate with conventional lithography techniques. Two-photon polymerization is a technique well suited for fabricating complex 3D geometries with sub-micrometer resolution. (**Right**) By mixing in diamond particles into TPP compatible resins, we aim to fix particles into a permanent, arbitrarily shaped structure.

We employ a two-photon polymerization (TPP) fabrication technique that provides new design flexibility for microscale applications. TPP is a 3D microfabrication process that utilizes high intensity, ultrashort laser pulses to initiate two photon absorption and the subsequent polymerization in photoresist.[26] TPP can create three-dimensional structures hundreds of microns in size with feature resolutions below 100 nm[27], complex freeform surfaces[28], delicate overhangs[29], and freely moving, independent components.[26,30] TPP is already widely employed for micro optics[28,31,32], microfluidics[33,34], metamaterials[35], and emerging applications such as microneedles.[36]

Through two-photon lithography, we demonstrate fabrication of arbitrarily shaped microstructures with diamond nanoparticles incorporated within them. Additionally, the microstructures can function as scaffolds for the placement of nanodiamond particles, enabling novel approaches to sensor arrangement. Quantum sensing of temperature and DC magnetic fields rely on optically detected magnetic resonance (ODMR) readout of the NV centers, accompanied by lock-in methods to mitigate autofluorescence from the structures and enhance sensitivity.

## Results and Discussion

In the following we demonstrate the fabrication of complex overhanging structures, verifying the incorporation of NV centers into the structures, and using a widefield imaging technique that can maximize optical contrast between NV center emission and the strong background photoluminescence from the TPP resin.

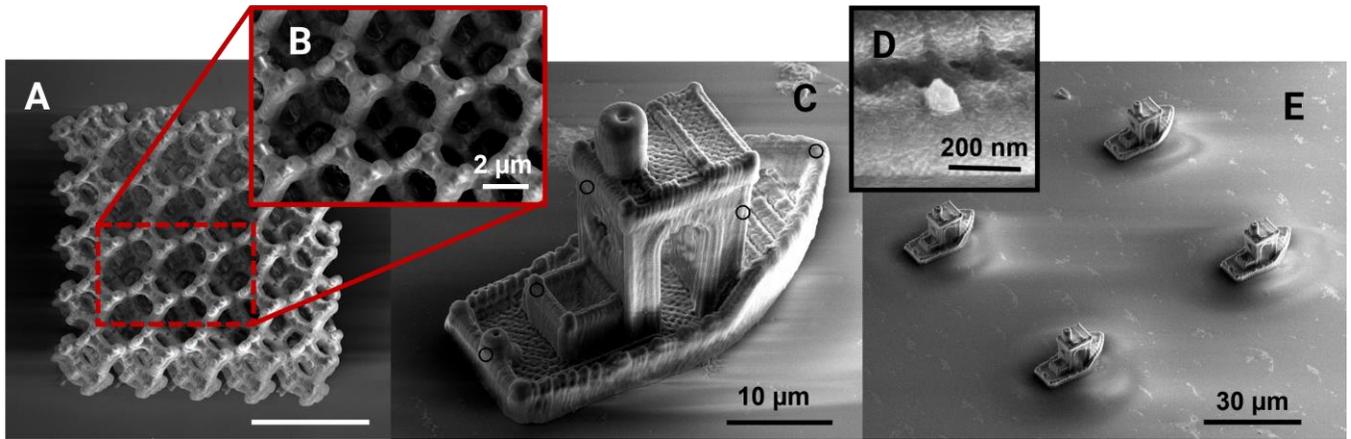

**Figure 2: Example TPP Structures** (A) SEM image of a porous tetrakaidecahedron structure exemplified in **Figure 1** fabricated in our setup with resin containing 25 nm diameter diamond NV center particles (scale bar 25 um) (B) with hollow interiors. (C) A microscale "Benchy" structure with submicron feature sizes patterned on a glass substrate (D) containing 100 nm diameter diamond particles where some are embedded onto the surface (circled) (E) These structures are arrayed to show the reproducibility and patterning capabilities of this technique with arrays of hundreds of elements being possible.

Structures are fabricated by employing sub-µm resolution direct femtosecond laser writing using two-photon polymerization (TPP) onto a biocompatible photoresist mixed with a concentrated solution of diamond nanoparticles. More details on the resin preparation and processing setup can be found in the **Material and Methods** section and **SI**. Our TPP processing setup (**Figure S1**) is capable of high-speed laser writing up to 6 mm/s. We readily achieve feature resolutions down to 400 nm. Using stimulated emission depletion techniques like Wollhofen et al., we expect to be able to reduce feature sizes below 100 nm.[27] Further discussions on print quality and comparison with the state of the art are elucidated in **Table S1**.

TPP is well suited for a variety of unmet applications in quantum sensing. For instance, TPP can construct complex porous structures and hollow microfluidic channels[37] as well as mesoscale objects[38] that can be patterned across large arrays without loss of feature resolution. **Figure 2** displays representative structures fabricated with resin containing 25 and 100 nm nanodiamonds. **Figure 2A&B** exemplifies one such porous structure with 2-5 µm pore size, suggesting potential applications in continuous-flow fluidic sensing. Given that TPP can be printed onto existing microfluidic chips, one could imagine using hybrid lithography processes to incorporate on-chip sensing elements for high-throughput assays.[39,40]

**Figure 2C&E** highlight the ability of TPP to generate finely detailed and complex geometries that can span hundreds or even thousands of micrometers. The depicted structure ("Benchy") is incorporated with NDs and arrayed over a larger area. We estimate the volumetric fraction of diamonds in the structures after development to be ~.2% based on ODMR images (see **Figure 4C**). For 100nm diameter particles this corresponds to ≈1 particle per 20 µm$^3$ and for 25 nm particles we estimate there to be approximately ≈1 particle per µm$^3$. These ratios are largely tunable based on the relative concentration of diamond solution to resin. Inevitably, given the high concentration of diamonds, several diamonds visibly appear on the surface of the structures in **Figure 2**. These are highlighted as circles in **Figure 2C**. **Figure 2D** depicts a closeup view of one of these 100 nm diamond particles lodged onto the surface of the parent structure. From the demonstrations in **Figure 2**, it is conceivable to fabricate micro lens arrays or photonic elements embedded with quantum sensors, enabling large field-of-view sensing applications.

**Optically Detected Magnetic Resonance**

Quantum sensing measurements utilizing ODMR rely on the spin-state-dependent fluorescence of NV center defects. **Figure 3A** presents a simplified model of an NV center in diamond, depicting it as a two-electron system with energy levels illustrated in **Figure 3B**. The ground triplet spin state exhibits zero field splitting (ZFS) at 2.87 GHz. The Hamiltonian of this system is described in the **SI**.

When microwaves (MWs) are applied resonant with the transitions between spin sub-levels, the optical emission intensity undergoes a change, providing a means to optically probe the energy levels through a sweep of MW frequency. In the presence of an external magnetic field, the degeneracy of the $|m_S = \pm 1\rangle$ magnetic spin sub-levels is lifted, while the ZFS value itself is temperature-dependent. These ODMR measurements can therefore be employed for tasks such as magnetic field or temperature sensing.

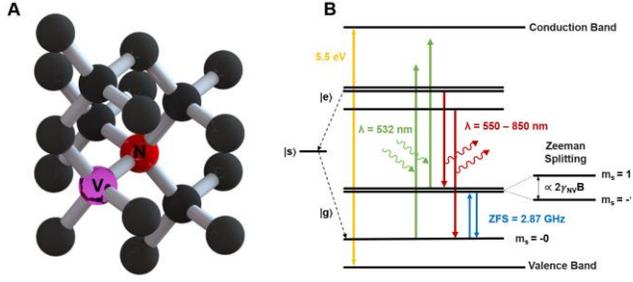

**Figure 3: Diamond NV Center Structure and Energy Level Diagram** (A) A nitrogen defect (red) is implanted adjacent to a vacant lattice site within a carbon bulk. In typical diamond nanoparticles, NV defects are present at concentrations ranging from 1 to 5 parts per million (ppm). (B) The fluorescence and spin polarization of the NV center depend on its spin relaxation dynamics. Upon optical excitation (green arrows), the non-radiative intersystem crossing to the singlet state exhibits a higher probability when excited to the $m_s= +1$ spin sublevel compared to the $m_s=0$ sublevel. This disparity enables the measurement of spin polarization through fluorescence intensity.

To understand the emissive characteristics of our functionalized TPP structures we first measure the photoluminescence spectra of both NV center containing nanodiamonds and the post-processed photoresist individually. The emission spectra under 532 nm excitation of the two materials (**Figure 4A**) show broad overlap between ~575 nm and 675 nm. Despite filtering our fluorescence readout to wavelengths >680 nm the optical signal is orders of magnitude larger than that of the NV centers based on the compositional fraction of the two species. To isolate the weaker emission of the NV centers from the background resin, we utilize a custom microscope (**Figure S2A**) capable of applying amplitude modulated microwaves across the NV center spin state transition frequencies range of interest (~2.800 – 2.925 GHz). Since only the NV center fluorescence intensity is modulated by the application of the microwaves within this range, the oscillating signal is extracted and amplified with a lock-in amplifier.

Initially, to verify the successful incorporation of diamond nanoparticles into our TPP structures we construct a widefield ODMR contrast image of a 15 μm diameter, 5 μm tall cylindrical structure, containing 100 nm diamond particles by applying microwaves at 2.87 GHz and subtracting out the background fluorescence (**Figure 4C**). This methodology is employed in a variety of other works on NV center sensing.[22,41–44] In this image the intensity of each camera pixel is determined by:

$$I_{contrast} = |I_{mw\,(on)} - I_{mw\,(off)}| \quad (1)$$

In these fluorescent contrast images (**Figure 4C**) we can clearly discern and locate at least 30 individual and possibly aggregated diamond particles within their fluorescent polymer matrix whereas these distinctions cannot be made in the corresponding fluorescent image of the pillars (**Figure 4B**). By using a more sensitive multi-pixel photon counter (MPPC) and sweeping across a range of microwave frequencies following the pulse sequence shown in **Figure S2B**, we construct the ODMR contrast spectrum shown in **Figure 4D** wherein percent contrast is defined as:

$$I'_{contrast} = \left(1 - \frac{I_{mw\,(on)}}{I_{mw\,(off)}}\right) * 100 \quad (2)$$

Subsequently we fit two exponentially modified gaussian functions to the data. At ambient conditions, we observe maximum ODMR contrasts at ~2.8616 GHz and ~2.8792 GHz. This low-field splitting is a remnant of residual strain in the diamond lattice after fabrication of nanoparticles. Using our NV-pillar structures we achieve a maximum ODMR contrast in the range of 1.2-2.5% for different structures likely based on the total number of NV centers randomly incorporated into each structure.

When varying the temperature of the diamond lattice in the NV molecular model theorized by Doherty et al.[45] the splitting parameter, D in the zero-field splitting term can be expressed as:

$$D \propto C\eta^2 \left\langle \frac{1}{r^3} - \frac{3z^2}{r^5} \right\rangle \quad (3)$$

Where C is the spin-spin interaction constant, $\eta$

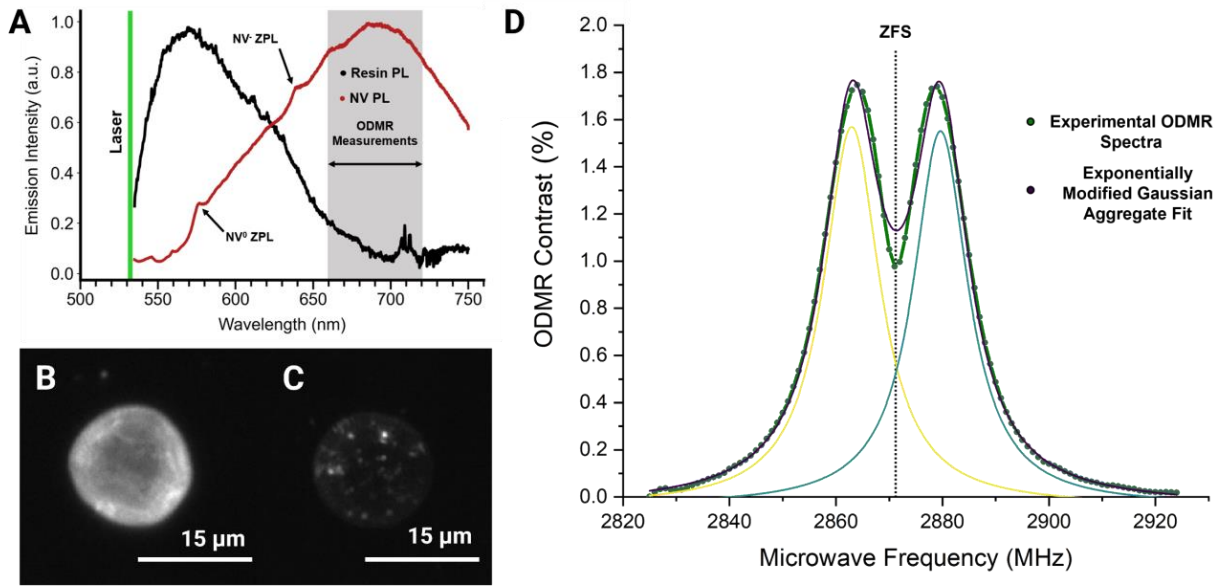

**Figure 4: ODMR Imaging** (A) Photoluminescent spectra of NV centers and SZ2080 resin as the TPP matrix. (B) Fluorescence image of a cylindrical TPP structure (C) in comparison to an ODMR fluorescent contrast image taken at 2870 MHz which clearly discerns the relative spatial location of NV center containing diamond nanoparticles. (D) ODMR contrast spectra, I'$_{contrast}$, of a pillar structure taken in ambient conditions.

the electron density, and $\left\langle \frac{1}{r^3} - \frac{3z^2}{r^5} \right\rangle$ the interaction of sp$^3$ electron densities in carbon atoms. As the temperature of the diamond nanoparticle increases, we expect thermal expansion of the lattice that increases the lattice spacing between atoms, r, resulting in a decrease in the splitting parameter.[45] In the ODMR spectra this manifests as a mild shift in the zero-field splitting to lower frequencies. Within relatively small ranges of temperature the change in the splitting parameter is locally linear.

In our experiments we investigate optical thermometry of 5 μm tall, 15 μm diameter cylindrical structures in temperature ranges varying from (295°K – 323°K) which are relevant to biological applications. Temperature is varied in these experiments by using a heating stage which is allowed to reach thermal equilibrium before collecting each temperature measurement. The resultant ODMR spectra are shown in **Figure 5A**. From these spectra we can calculate the ZFS and uncertainty by fitting a two-peak exponentially modified gaussian function to each spectrum, the results of which are shown in **Figure 5B**. We observe a strong linear relationship of -.07447 ±.00154 MHz/°K which agrees with measurements from Doherty et al.[45] and achieve an average temperature sensitivity of .645 K/$\sqrt{Hz}$. Details on the sensitivity calculations are provided in the **SI**.

Lock-in ODMR techniques excel at improving the sensitivity of temperature measurements and the structuring enabled by TPP provides capabilities that are not achievable for other methods of NV based thermometry. For example, in prototypical biological applications NV centers are disbursed into solution and local temperature measurements of cells and organisms are dependent on intercellular uptake of particles. Due to diffusion in the cellular fluid medium, sensing particles would not maintain a set distance from the observables of interest. Incorporating the particles into scaffold members could provide extracellular support structure onto which the cells attach that simultaneously acts as a medium for temperature and magnetic field sensing. More traditional ODMR techniques likewise suffer from the background autofluorescence of their surroundings. Stray organic molecules and thermal noise can contribute fluorescence at relevant photon energies for ODMR that would also be modulated with techniques that use laser pulse repetition frequency as the lock-in frequency. Since only the NV centers fluorescence is modulated within the range of relevant microwave

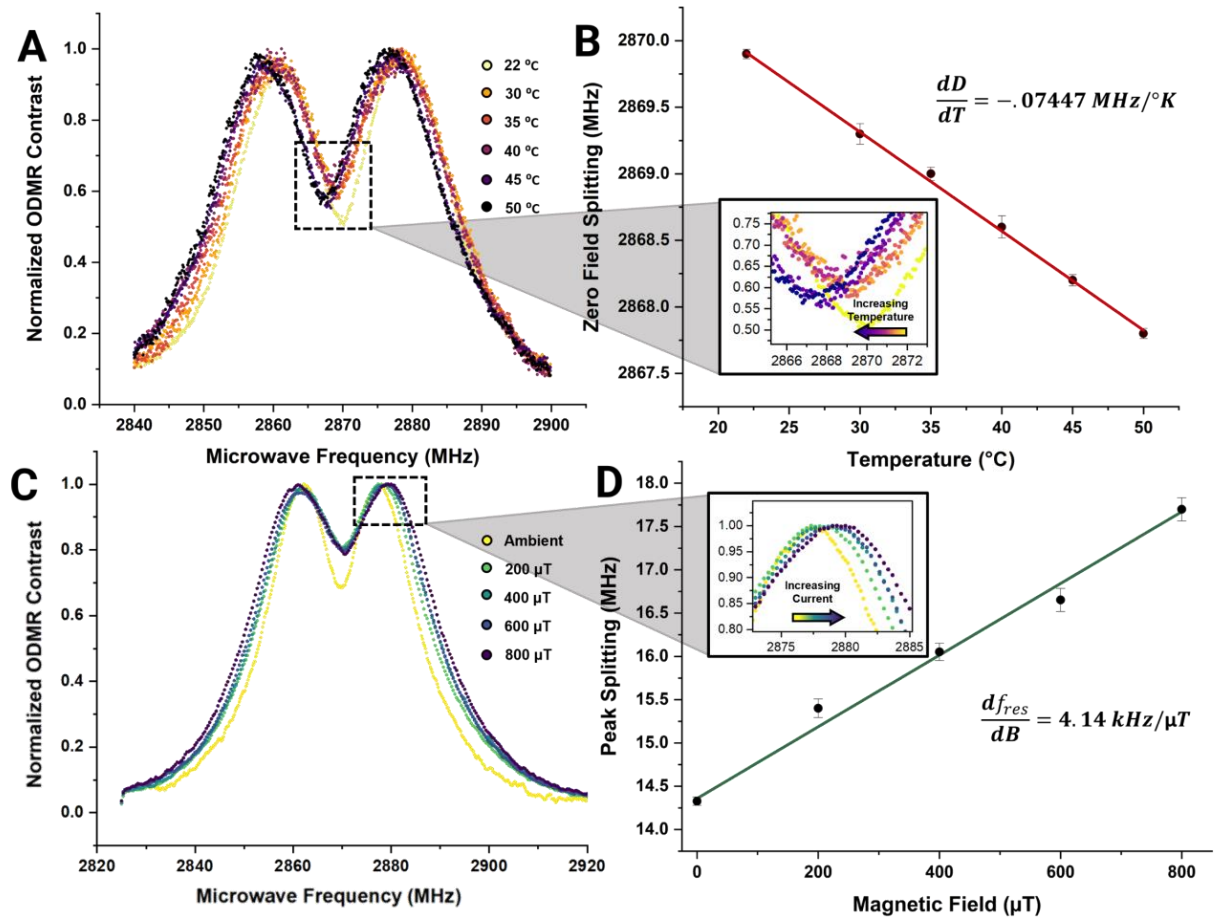

**Figure 5: Quantum Sensing Measurements** (A) ODMR contrast spectra of NV center containing microstructures normalized to maximum contrast at varying temperatures (22-50°C) relevant to bio-thermometric applications (B) Zero-field splitting in ODMR as a function of changing temperature where a clear locally linear relationship of $-.07447\ MHz/°K$ is observed. Error bars are 95% confidence intervals. (C) ODMR contrast spectra of sensing structures at varying applied magnetic fields

frequencies, we can effectively isolate the ODMR signal and measure a maximum observed contrast signal to >2% with high signal-to noise ratio which is on par with measurements observed elsewhere in literature.[7]

In an ensemble of nanodiamonds, the quantization axis for the NV centers, aligned with the N-to-V lattice axis, is randomly oriented. Consequently, NV centers within each particle encounter a distinct magnetic field projection along this direction compared to NV centers in other particles within the resin matrix. In the collected ODMR spectra this is manifested as a mild increase in the splitting of the two peaks as well as a broadening of the individual peaks. This broadening is explained in greater detail in **Figure S3.**

We perform a demonstrative experiment by passing a variable current through a thin wire placed roughly 50 μm away from a stationary cylindrical structure. We collect ODMR measurements at the same distance from the wire for each current. We estimate the magnetic field by using an infinite wire approximation,

The results of this experiment are shown In **Figure 5C**, where discernable splitting effects are observed. By fitting an exponentially modified gaussian functions to this data we calculate the splitting between peaks of the constituent curves as well as an associated uncertainty and plot the results in **Figure 5D**. A linear fit yields a relationship of 4.14 ± .209 kHz/μT for this experimental system. We find an average sensitivity of approximately 216.4 μT/$\sqrt{Hz}$. using a similar estimation procedure as in the temperature measurements. We anticipate that more advanced protocols, e.g. employing a Ramsey se-quences, can be employed to yield better measurement sensitivity.[46,47]

**Outlook**

The ability to incorporate diamond particles containing NV centers into our TPP resin gives extraordinary design freedoms for building three dimensional structures with microscale features that can be patterned or arrayed across a substrate or device. This simultaneously enables sensing applications for point mapping across large areas as well as 3D mapping of microscale volumes. Herein we have demonstrated the feasibility of measuring ODMR spectra within single structures by using lock-in amplification to isolate NV center emission from fluorescent background. We leave room for further work to apply these techniques to demonstrate large field 2D and microscale 3D imaging for application specific experiments.

TPP has already found use in a number of applications discussed previously. Embedding nanodiamonds into structures for these applications would allow for enhanced diagnostic capabilities that could add new dimensionalities to the possible *in-situ* measurements of these systems. We can also envision a number of new applications that are discussed in the **Supplementary Information (SI)** including functionalized cellular scaffolding, remote detection of passing currents on microelectronics, and chemical sensors in microfluidics.

While this work focuses on optical detection of magnetic resonance of NV center particles, it should be noted that optical detection of magnetic resonance can occur in other point defect systems such as silicon vacancy SiC, hBN, and silicon at cryogenic temperatures.[3,5,48] Similar techniques for incorporating color center particles into TPP resins could allow for tunable sensing properties.

**Conclusion**

We have demonstrated a technique for developing "designer" microscale 3D structures relevant for quantum sensing by mixing nanoparticles containing NV centers into TPP compatible resins. We observe that autofluorescence from the bulk resin matrix is more than an order of magnitude larger than the ODMR signal and thus construct a differential contrast microscope using lock-in amplification at the applied microwave amplitude pulse frequency to isolate ODMR signal. Finally, we show that combining these two techniques allows us to measure temperature and magnetic field measurements. Further work should focus on the development of analogous 3D imaging techniques using confocal microscopy and exploring practical end applications where this technique would be advantageous.

**Acknowledgements**

Brian Blankenship acknowledges support from the NSF Graduate Research Fellowship (DGE 2146752). Support to the Laser Thermal Laboratory by the National Science Foundation under grant CMMI-2124826 is gratefully acknowledged. We acknowledge support from DOE BER award (DE-SC0023065). This work was partially funded by the Laboratory Directed Research & Development program at the E.O. Lawrence Berkeley National Laboratory. SEM images were taken with the Scios 2 DualBeam available at the Biomolecular Nanotechnology Center of the California Institute for Quantitative Biosciences, UC Berkeley.

**Materials and Methods:**

**Resin Preparation:**

A hybrid organic-inorganic resin, SZ2080 is used with Zr-DMAEMA (30wt%) as a binder. The resin is composed of 70 wt% zirconium propoxide and 10 wt% (2-dimethylaminoethyl) methacrylate (DMAEMA) (Sigma-Aldrich). A suspension of either 25 or 100 nm diameter NV center diamond particles in water (Adamas Nanotechnologies) was included in the resin at a ratio of 1:50 (v/v) and mixed for 30 minutes.

**Structure Fabrication:**

Structures were fabricated by sub-micron resolution direct femtosecond laser writing using two-photon polymerization on SZ2080 photoresist. The source laser is a FemtoFiber Pro NIR laser which emits 780 nm, 100 fs FWHM, pulses at 80 MHz. Polymerization of the resin is achieved with a 40x microscope objective lens (Plan-Apochromat 40×/1.3 Oil Olympus). The laser output energy was measured before the objective lens at 6.6 mW, and the scanning speed of the laser spot at the imaging plane was set to 1000 μm/s and is steered by using a two-axis optical galvanometer. The resin sample is positioned with three axis piezo and servo stages. Structures are built by building successive layers in the optical axis by scanning the laser across the imaging plane at set heights.

**Photoluminescence Measurements:**

Photoluminescence measurements were taken with a Renishaw InVia spectrometer with 1800 lines/mm grating.

**ODMR Measurements:**

ODMR is measured using a custom built wide-field fluorescence microscope discussed in detail in **Supplemental Information**. A diagram of the microscope is shown in **Figure S2**.

**Supplementary Information**

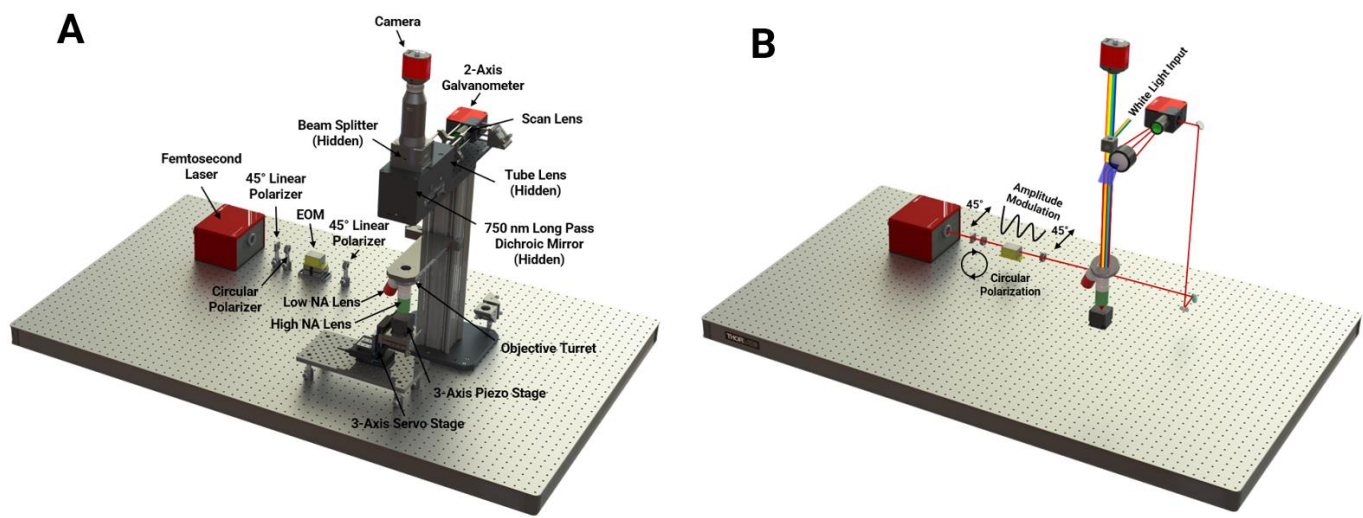

**Figure S1: Two-Photon Polymerization Setup** (A) Rendering of the TPP processing setup with adjustable power modulation and laser scanning capabilities. B) A circularly polarized femtosecond laser beams phase is modulated by an EOM and linear polarizer at the output. This beam is scanned by a galvanometer for fast (> 1000 μm/s) scanning at the image plane. A beam splitter allows a separate imaging light to image while printing.

**Table S1: TPP Print Quality and NV Composition in Resin**

|  | Minimum feature resolution | Volume printed per second | Maximum geometry constraints | Estimated particles per unit volume | NV centers per particle |
|---|---|---|---|---|---|
| Present Work | ~400 nm | 2,000 - 10,000 μm$^3$/s | 200 μm (xy) 170 μm (z) | ~.05 μm$^{-3}$ | ~2200 |
| Current state of the art | < 55 nm[1,2] | ~1,000,000 μm$^3$/s | >1mm (xy)[3] > 450 μm (z)[4] | N/A | N/A |

In the present work we build structures using a custom TPP setup capable of achieving 500nm feature sizes with a 1.3 NA objective at laser powers and galvanometer speeds specified in the paper. Using state of the art techniques such as stimulated emission depletion lithography (STED), TPP can construct objects with sub-55 nanometer feature resolution. Varying galvanometer speeds allows us to achieve print rates in the range of 2,000 - 10,000 μm$^3$/s whereas state of the art projection-type printers can reach print rates approaching 1,000,000 μm$^3$/s with micron-level resolution.

We estimate that the density of particles in our TPP structures is approximately 1 particle per 20 μm$^3$. This is controllable to some degree based on the ratio of diamonds to resin used in the printing medium. For our sensing measurements we utilize 100nm diameter NV center particles with 3 ppm NV centers from Adamas Nanotechnologies and estimate there to be 2200 NV centers per diamond particle.
12

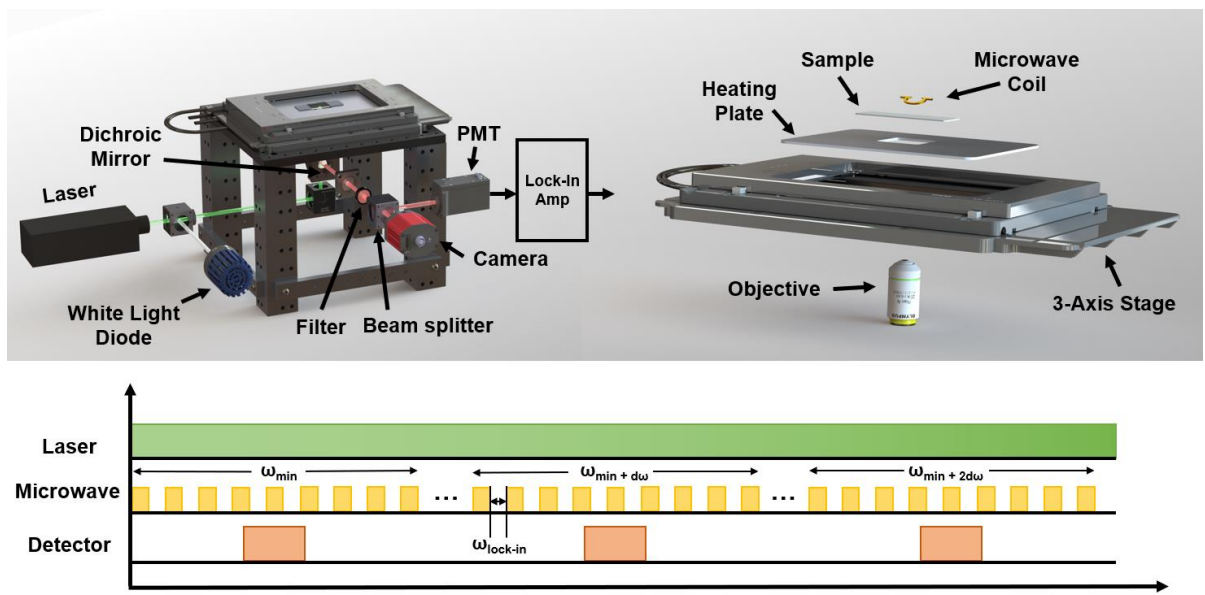

**Figure S2: ODMR Microscope and Pulse Sequence** (A) Left: Rendering of the ODMR microscope and beam path. A 532 nm excitation laser (Coherent Verdi G) and white light imaging source (CoolLED PE-4000) are directed to an objective (Olympus LMPLFLN 20X) via a dichroic mirror (Thorlabs FEL0600-1) and focused onto the sample. Lower energy photons from sample emission pass through the 600 nm longpass dichroic mirror and a 680 nm bandpass filter (FF01-680/42-25) to a tube lens. A beamsplitter directs approximately 30% of the light to an imaging sCMOS camera (Teledyne Kinetix) for imaging and 70% to a multi-pixel photon counter (Hamamatsu C14455) for ODMR measurements. The MPPC signal is amplified using a lock-in amplifier (SRS SR830) with reference frequency set to the microwave pulse frequency. Microwave excitation is generated using a synthesizer (HP 8664A) and 100% amplitude modulated using a switch (minicircuits ZYSW-2-50DR, Empower 1146-BBM4A6AK5). This modulating signal is then amplified by a series of amplifiers (minicircuits ZYSW-2-50DR) to a final power of approximately 50 W) Right: Closeup of the stage. A heating plate controls the temperature of the sample while a small microwave coil applies pulsed microwaves. (B) ODMR pulse sequence. A 100mW CW laser continuously excites the sample while a variable frequency microwave at constant pulse frequency irradiates the sample. The MPPC captures signal during both "on" and "off" phases of the microwave pulse.

## Exponentially Modified Gaussian Function Fitting

Our data is fit to exponentially modified gaussian functions of form:

$$f(x; \mu, \sigma, \lambda) = \frac{\lambda}{2} e^{\frac{\lambda}{2}(2\mu + \lambda\sigma^2 - 2x)} erfc\left(\frac{\mu + \lambda\sigma^2 - x}{\sqrt{2}\sigma}\right)$$

Where erfc is the complementary error function defined as:

$$erfc(x) = 1 - erf(x) = \frac{2}{\sqrt{\pi}} \int_x^\infty e^{-t^2} dt$$

The exponentially modified gaussian function is the convolution of the normal and exponential probability density functions and was found to be the best fit for our ODMR contrast data.

## NV Center Hamiltonian

The Hamiltonian of an NV center system can be described by:

$$H_{NV} = D\left(S_z^2 - \frac{1}{3}S^2\right) + \gamma_e B_0 (S_x \sin\theta + S_z \cos\theta) + SAI$$

Where D is the zero-field splitting (ZFS) parameter, $\gamma_e = 28\ MHz/mT$ is the electron gyromagnetic ratio, S represents spin angular momentum, B is the magnetic field, A the hyperfine tensor, I the spin operator of the nitrogen nucleus and $\theta$ is the angle between the magnetic field and NV axis. By precisely measuring the spectral emissions governed by the above Hamiltonian we measure the local temperature and magnetic field in our structures.

Changes in the measured ODMR spectrum can be used as a proxy to measure the local conditions of the ensemble of NV centers inside each structure. The Hamiltonian in equation (1) can be understood as the sum of the zero-field splitting, electron Zeeman term, and the hyperfine coupling:

$$H = H_{ZFS} + H_{Zeeman} + H_{HF}$$

Where differences in temperature more strongly affect the contribution of the zero-field splitting term in the Hamiltonian, and the Zeeman term is largely influenced by the magnetic field effectively allowing a means to decouple measurements of magnetic field and temperature.

## Sensitivity Calculations

Sensitivity was calculated in the following way:

$$\text{Sensitivity } \eta = \frac{\sigma_P \sqrt{\delta t_{int}}}{\frac{dX}{dY}}$$

Where $\sigma_P$ is the standard error of a point measurement in units of MHz, $\delta t_{int}$ is the integration time of the point, and $\frac{dX}{dY}$ is the change in position X in units of MHz with respect to parameter Y (temperature, magnetic field).

### Figure S3: Magnetic Field Broadening Simulations

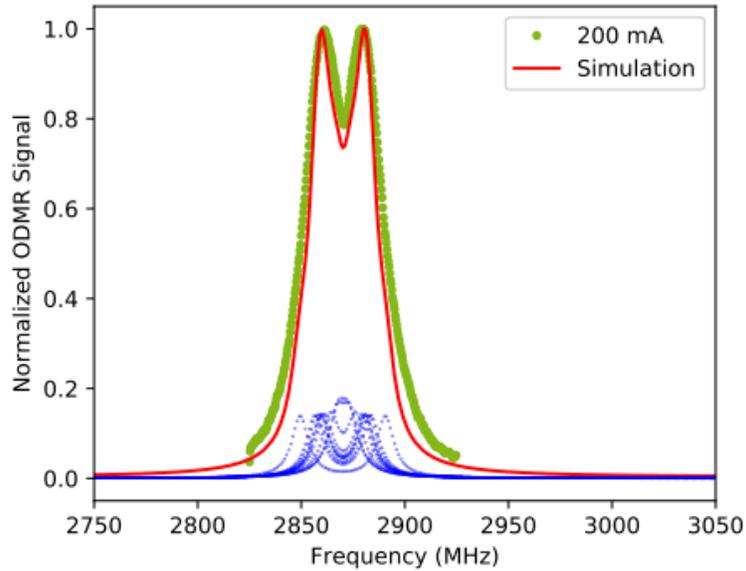

We performed simulations of an ensemble of randomly oriented NV centers in order to understand why we observed broadening of peaks in the presence of magnetic fields. In the graph, the red curve is the sum of the ODMR signal of ten randomly oriented NV centers simulated with the Lindblad equation. The Lindblad equation is integrated for half a microsecond, and the ODMR signal is calculated from the resulting density matrix. The Range-Kutta method is used for numerical integration. A similar methodology is employed in Singh et al.[5]

**Table S2: Outlook of TPP Quantum Sensing**

| Field | Application | Challenges | Solutions |
|---|---|---|---|
| **Biology** | 2/3D Bio-Thermometry 3D Bio-Thermometry | Diffusion of particles in-vivo prevents imaging of NV centers at a fixed location. | Construction of 2D cellular scaffolds (see Trautmen et. al.[6]) with nanodiamonds integrated into the resin matrix for fixed-position imaging. |
| | Neuronal Imaging | Photobleaching of fluorophores over prolonged exposures. | Construction of 2D/3D cellular scaffolds (See Accardo et. al.[7]) with nanodiamonds integrated into the resin matrix for fixed-position imaging. NV centers are also not photobleachable. |
| **Semiconductors** | Remote Detection of Passing Currents | Commercial magnetometers cannot be used for sub-millimeter detection volumes and NV centers cannot ordinarily be patterned onto non-diamond substrates. | Structures containing NV centers can be patterned onto most substrates and can be localized to microscale detection domains. |
| | Widefield Localized Surface Temperature Measurements | Surface thermometry techniques such as TDTR require addition of thin films onto substrate, thermocouples are bulky and require wired connections. | Structures containing NV centers can be printed onto most substrates, can be localized to microscale detection domains, requires no wires, and measurements are independent of optical properties of substrates/thin films. |
| **Microfluidics** | Chemical Sensing | Microfluidic channels are printed onto costly diamond substrates. | Porous structures can be directly printed into microfluidic cavities with sub 60 nm resolution feature sizes for micro/nanoliter confined volume NMR. |
| | Flow Temperature Measurements | Microfluidic channels must be printed onto costly diamond substrates or incorporate diamond particles into flow. Anemometers dependent on fluid/flow properties. | Fixed structures along a microfluidic channel can be constructed to monitor temperature at fixed locations. Measurements are independent of fluid/flow properties. |